\documentclass[a4paper,11pt]{article}

\usepackage{jinstpub} 
\usepackage{lineno}
\usepackage{titlesec}

\title{\boldmath Studies of time resolution, light yield, and crosstalk using SiPM-on-tile calorimetry for the future Electron-Ion Collider}

\author[a,b]{Miguel Arratia,}
\author[a]{Luis Garabito Ruiz,}
\author[a]{Jiajun Huang,}
\author[a]{Sebouh J. Paul,}
\author[a]{Sean Preins,}
\author[a]{Miguel Rodriguez}

\affiliation[a]{Department of Physics and Astronomy, University of California, Riverside, CA 92521, USA}
\affiliation[b]{Thomas Jefferson National Accelerator Facility, Newport News, Virginia 23606, USA}

\emailAdd{miguel.arratia@ucr.edu}

\abstract{We recently proposed a high-granularity calorimeter insert for the Electron-Ion Collider (EIC) that is based on plastic scintillator tiles readout with silicon photomultipliers. In this work, we concretize its design by characterizing its building blocks with measurements of light yield, optical crosstalk, and timing resolutions using cosmic-rays, an LED, and a beta source. We also compared two approaches for the optical isolation of cells: ``megatiles'' with grooved boundaries between cells, and a 3D-printed plastic frame hosting individual cells. We found that the latter suppresses optical crosstalk to negligible levels while providing an easier assembly method. Overall, these performance studies can help inform calorimeter design and realistic simulations of 5D showers (time, energy, position) for the EIC and other experiments.
}

\keywords{Calorimeters; Scintillators and scintillating fibres and light guides; Scintillators, scintillation and light emission processes (solid, gas and liquid scintillators); Detector design and construction technologies and materials;}

\begin{document}

\maketitle
\flushbottom

\section{Introduction}
Highly granular sampling calorimeters based on scintillators readout directly with silicon photo-multipliers, also known as a ``SiPM-on-tile'' approach~\cite{Blazey:2009zz,Simon:2010hf}, have been proposed for a number of experiments in various facilities~\cite{CALICE:2010fpb,CALICE:2022uwn,Dannheim:2019rcr,CEPCStudyGroup:2018ghi,JIANG2020164481,Li_2021,Jiang:2020tve,Duan_2022,ILDConceptGroup:2020sfq} to enable the particle-flow paradigm~\cite{Thomson:2009rp,RevModPhys.88.015003}; notably, a large-scale use is being deployed for an upgrade for CMS at the LHC~\cite{CMS:2017jpq}.  
A large number of beam-test studies of this technology have been performed by the CALICE Collaboration~\cite{RevModPhys.88.015003}.

The SiPM-on-tile approach provides flexibility in calorimeter design; for example, the shape and size of cells can be straightforwardly tuned, varying with position in the detector for specific needs based on factors such as cost, noise, or radiation tolerance.  It also requires a less arduous and complex assembly procedure compared to traditional designs based on waveshifting fibers. 

We recently proposed to use the SiPM-on-tile  approach for a design of a calorimeter insert for the ePIC detector at the future Electron-Ion Collider (EIC)~\cite{hcalInsert}, which would cover $3<\eta<4$ and maximize acceptance near the beampipe. Each layer in the calorimeter is slightly different to accommodate a beampipe with a 25~mrad crossing angle. As such, the natural choice for the shape of most cells is hexagons, which can help tessellate complicated areas effectively. Moreover, the design is enabled by the fact that the EIC generates relatively minor radiation fluence---not exceeding $10^{12}$~neutrons/cm$^{2}$ per year at design luminosity $(10^{34}$~cm$^{-2}$s$^{-1} )$~\cite{AbdulKhalek:2021gbh}.  The modest amount of radiation damage to SiPMs could be mitigated with annealing after each run~\cite{Garutti:2018hfu,hcalInsert}.

The design for the calorimeter insert is driven by the need to maintain a sufficient signal-to-noise ratio at EIC radiation fluences without active cooling, large light yield to keep the possibility of cell-by-cell calibration with minimum-ionizing particles, as well as the need to maximize acceptance in the complicated volume near the EIC beampipe. The area of the calorimeter insert is only about 60$\times$60~cm$^{2}$, so constraints related to construction at scale and cost are not as stringent.

In this work, we seek to define the parameters of the building blocks suitable for the EIC application, specifically the light yield, time resolution, and optical crosstalk. We performed studies using UV LEDs, cosmic rays, and a radioactive source (Sr-90).
 We also determined the level of optical crosstalk between cells using the traditional megatile approach and a novel strategy based on a 3D-printed frame.

\section{The high-granularity calorimeter insert}
The high-granularity calorimeter insert (HG-CALI) proposed in  Ref.~\cite{hcalInsert} was designed to maximize acceptance ``as much as technically possible'' to meet one of the key requirements of the EIC \cite{AbdulKhalek:2021gbh}.  At high pseudorapidity, high granularity is needed in order to achieve reasonable angular resolution and to disentangle nearby showers.  Further, the tracking performance is expected to degrade at small angles because of the solenoidal nature of the magnet to be used---making hadronic calorimetry crucial in this region~\cite{Arrington:2021yeb,AbdulKhalek:2021gbh}.

\begin{figure}[h!]
    \centering
    \includegraphics[width=0.99\textwidth]{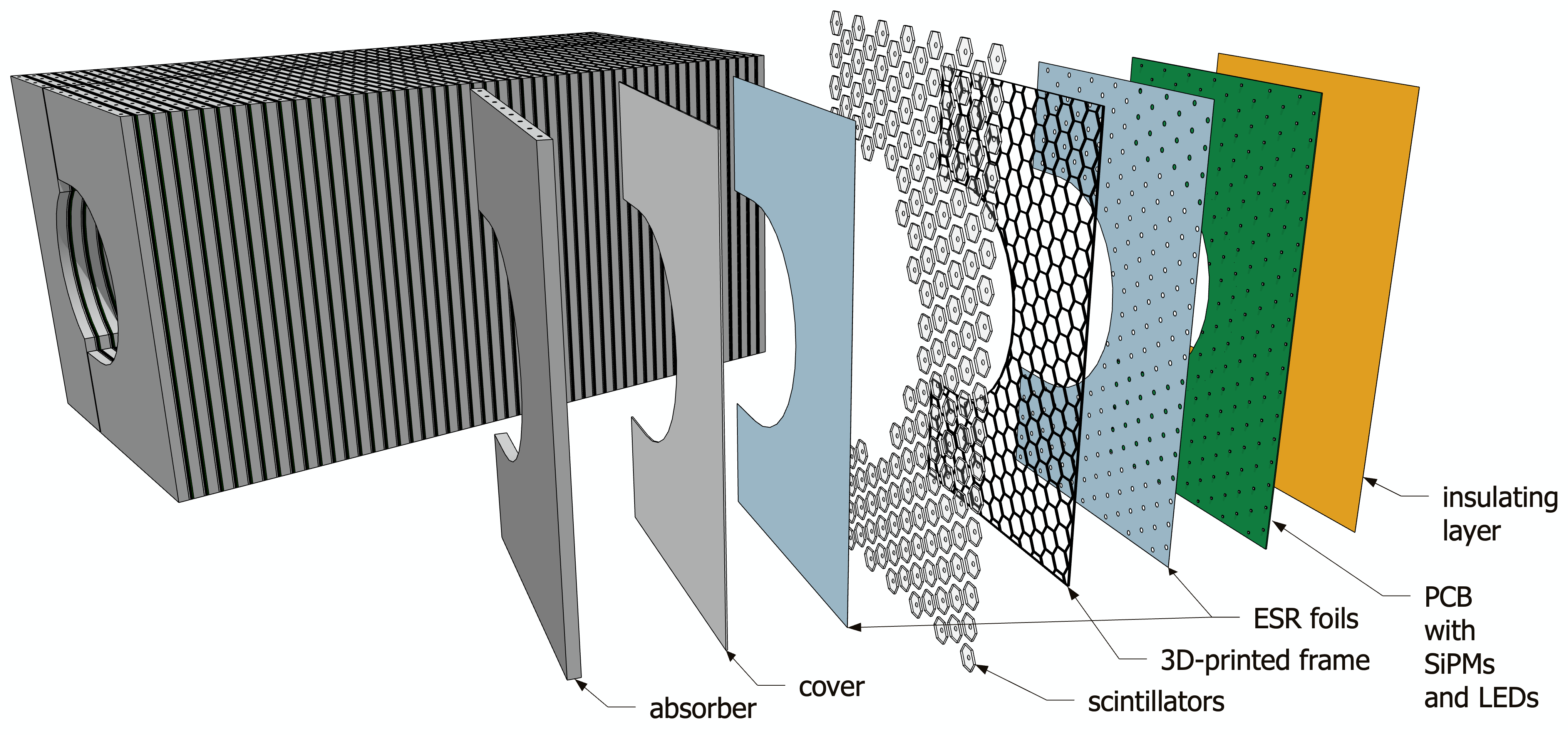}
    \caption{An exploded view of the high-granularity calorimeter insert proposed in Ref.~\cite{hcalInsert}, modified to use the 3D-printed frame strategy rather than the megatile. 
    }
    \label{fig:explode_view}
\end{figure}

Figure~\ref{fig:explode_view} illustrates the design that consists of alternating layers of absorber material (tungsten or steel, depending on the layer), and scintillators.  The scintillators are sandwiched between a cover and a SiPM-carrying printed circuit board (PCB), which includes an LED for each cell and temperature sensors for calibration and monitoring purposes. The detector as a whole is split into two parts that could be retracted laterally along with the two halves of the larger hadronic calorimeter endcaps, which will rest on a support structure with a rail system~\cite{AbdulKhalek:2021gbh}.

Two strategies are considered in this work for dividing the scintillator layers into cells.  The first strategy, which is used in the baseline design of the HG-CALI~\cite{hcalInsert}, is to use ``megatiles'', which are large pieces of scintillator divided into smaller readout cells by cutting grooves into them. This strategy has been employed, \textit{e.g.}, in the STAR~\cite{STAR:2002ymp} and CDF~\cite{DeBarbaro:267775} experiments.  It is also being considered by the CALICE collaboration~\cite{Laudrain_2022} as an alternative to using individual scintillators wrapped in reflective foil glued directly onto the PCB~\cite{Seflow2019}.      

The second strategy, which we propose and validate in this work, is to produce each cell separately, and then fit them into an opaque 3D-printed frame.  Instead of wrapping these cells in foil, we would use a single sheet of foil upstream and downstream of the frame, and use reflective paint on the edges of the tiles.
This makes the construction process easier than the delicate process of cutting the grooves in a megatile, while simultaneously eliminating the need to bend the foils.
\section{Experimental Setup for Prototyping Studies}
\subsection{Data Acquisition and Processing}
For our measured signals, we used 3x3~mm$^{2}$ SiPMs from Hamamatsu (model S14160-3015, which has about 40k pixels), mounted on a carrying board that provided bias voltage and a 20~dB voltage amplification~\cite{1861126}. We also used SenSL 6x6~mm$^{2}$ SiPMs (model MicroFJ-60035-TSV) in their evaluation boards for triggering. The SiPM signals were further amplified using one MiniCircuit ZX60-43-S+ before being readout by a DRS4 digitizer controlled with the RCDAQ software~\cite{RCDAQ}. The DRS4 board was set to a sampling rate of 2~GS/s, except in the timing-resolution setup where it was increased to 5~GS/s.

Depending on the setup, the trigger to record an event would be generated either by the trigger output of a LED Driver (CAEN SP5601) or from coincident signals generated by SiPMs affixed to scintillating tiles with a trigger logic implemented in the DRS4 digitizer.

The signals for each SiPM were calibrated using low-intensity LED pulses; a Fourier analysis of the observed photoelectron peaks yielded a gain factor of 6.3$\pm$0.1~mV/pixel consistently for the two SiPMs used for testing when biased at 40.5~V. This procedure was repeated varying the SiPM bias voltage; a linear fit to the gain vs.~bias voltage data was used to define the breakdown voltage as the voltage with zero gain. The breakdown voltage was found to be 38.3$\pm$0.6~V, with the two SiPMs analyzed yielding consistent results. The SiPMs were operated at 40.5~V, so slightly above 2~V overvoltage~\footnote{While this value is rather small, we used it for reference since it is the operating value used for the STAR HCAL, which uses SiPM readout under a similar radiation fluence to what the calorimeter insert will receive at the EIC.}. 

The pedestal noise of the setup was measured using randomly triggered data. It is well described with a Gaussian with width of 2 mV. The minimum signal recorded in either the measurement or trigger SiPMs was required to be at least 7 times this value, or about 2 photoelectrons.

The DRS4 digitizer records SiPM waveforms that typically consist of three parts: a background baseline, an avalanche rising peak, and an exponential decay. We implemented various algorithms to extract the pulse-height amplitude and the time of arrival of the pulse, which we call ``base'', ``smooth'', and ``fit''. In the base method, the pulse height is the difference between the maximum voltage in the waveform and the average of the pre-trigger baseline; then, it finds where the points behind the maximum value crosses the half value of the pulse-height to determine the time of arrival. The smooth method is similar to the base method, but with a five-midpoint averaging of the whole waveform. The fit method consists of a $\chi^{2}$ fit to the whole waveform into three different segments: horizontal, linear, and exponential decay. The pulse-height value is determined as the difference between the peak value of the linear segment and the baseline value; the time of arrival is the same as in the previous method. 

We perform a quality control procedure on the measured waveforms, which some times exhibit nonphysical spikes. After identifying such spikes, we replace the sample values in the spike with values obtained with a five-midpoint smoothing. We removed poor-quality events (caused by pileup or cross-triggering within the SiPM) using a cut on the $\chi^2$ obtained with the fit method.

\subsection{Prototype scintillators}
\begin{figure}[h!]
\centering 
\includegraphics[width=0.99\textwidth]{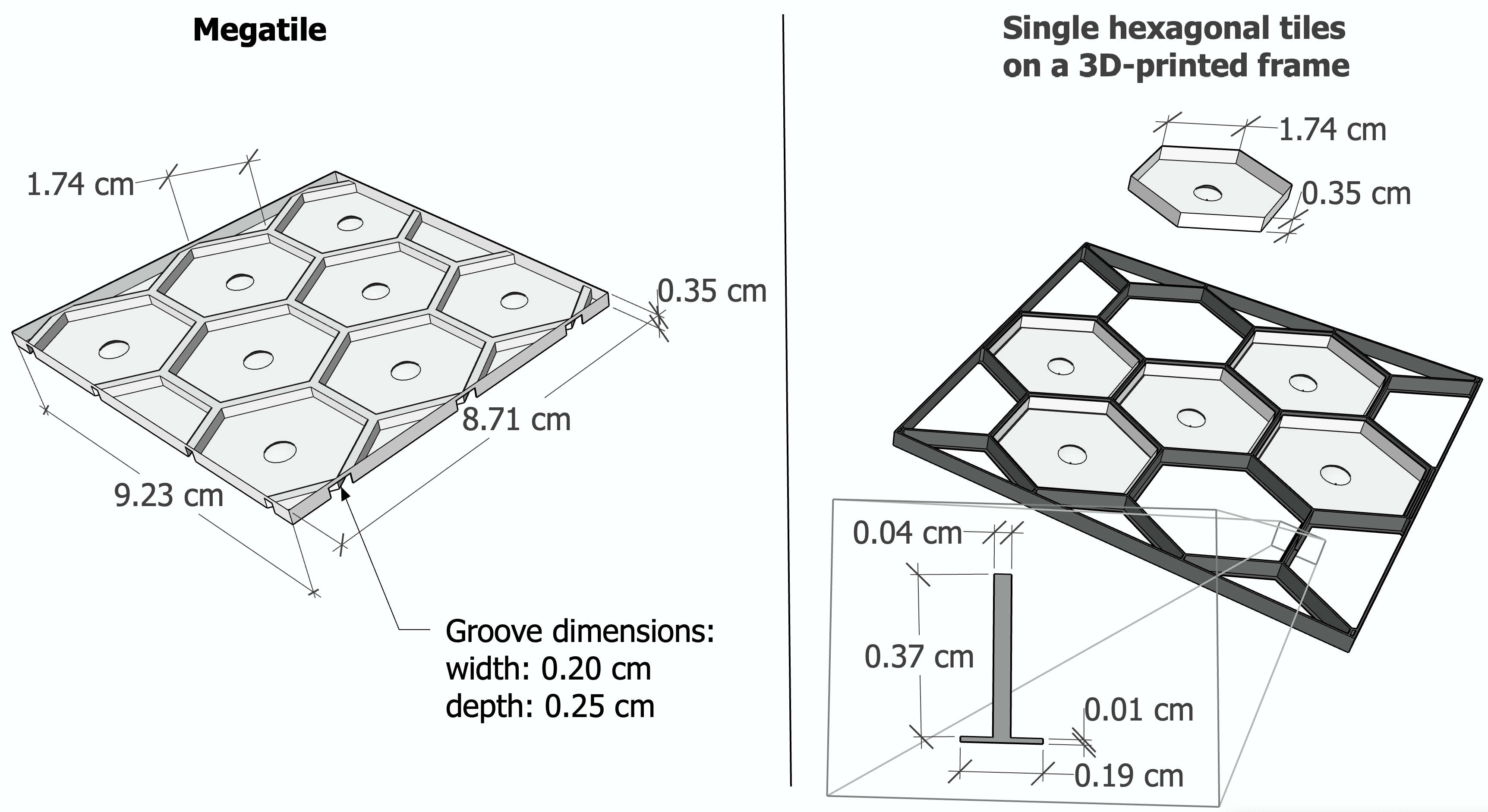}
\caption{\label{fig:scintillators} Renderings of the megatile (left) a single hexagonal tile (upper right), and a 3D-printed frame (bottom right).  An inset shows a cross-sectional view of the frame.  
}
\end{figure}

We used two types of scintillator tiles in this study.  The first type was a megatile with transverse dimensions 9.2$\times$8.7~cm$^2$ and a thickness of 3.5~mm.  This was divided into hexagonal cells (area 7.9 cm$^2$, side length $\approx$1.7~cm) via grooves (2.5~mm deep and 2~mm wide).  The inside of these grooves were painted with reflective white paint (Saint-Gobain BC-621), similar to the approach used in the CMS hadronic calorimeter~\cite{CERN-LHCC-97-031}.  The second type of tile was a hexagon with the same size and thickness as one of the cells of the megatile. In both cases, the cells have a polished dimple in the center to fit an SiPM to improve signal uniformity across the area of the cell~\cite{Blazey:2009zz,Simon:2010hf}. Renderings of the design for these prototypes are shown in Fig.~\ref{fig:scintillators}.  

The thickness of the scintillator tiles used in these studies (3.5~mm), as well as the hexagonal cell area (7.9~cm$^2$), are comparable to those of the HG-CALI design proposed in Ref~\cite{hcalInsert}. The dimple at the center of each cell has the geometry of a sphere of radius 3.8~mm going 1.6~mm deep within the tile, following  Ref.~\cite{Belloni_2021}.   
 The composition of the scintillator tiles is EJ-212, which is a type of polyvinyl toluene- (PVT-) based plastic that is chosen because of its radiation hardness~\cite{Liao_2015} and light yield~\cite{Chatrchyan_2018}.

In addition to the prototype megatile, we also created a prototype 3D-printed frame of black PLA, as shown on the right side of Fig.~\ref{fig:scintillators}.  The printer we used was an Ender 3 V2 using a 0.4~mm nozzle. In this frame, a 0.4~mm thick wall separates the individual cells, allowing for a dead zone between cells that is 1/5 as wide as the one in the megatile setup.  Thin shelves on one side of the frame are used to hold the scintillator tiles in place.  However, these shelves were only included for prototyping studies and we do not plan to include them in the final design of the HG-CALI.

For both the megatile and the individual hexagonal-cell tiles, the top and bottom of tiles were covered with 3M Vikuiti ESR foil, and the edges were painted with the reflective paint. This was done to increase light yield and improve uniformity~\cite{Blazey:2009zz,Simon:2010hf,Abu-Ajamieh:2011mao,Belloni_2021}. The ESR foil was chosen over alternatives such as Tyvek due its better performance as demonstrated, \textit{e.g.}, in Ref.~\cite{Belloni_2021}.  

Using the paint on the edges, and using foil only on the top and bottom of the tile (rather than fully wrapping the tile in foil like in the CALICE prototypes~\cite{Seflow2019}) simplified the production process, since the foil is difficult to bend and doing so can create air gaps.  

\subsection{Cosmic-ray setup}
\begin{figure}[h!]
\centering 
\includegraphics[width=0.49\textwidth,clip]{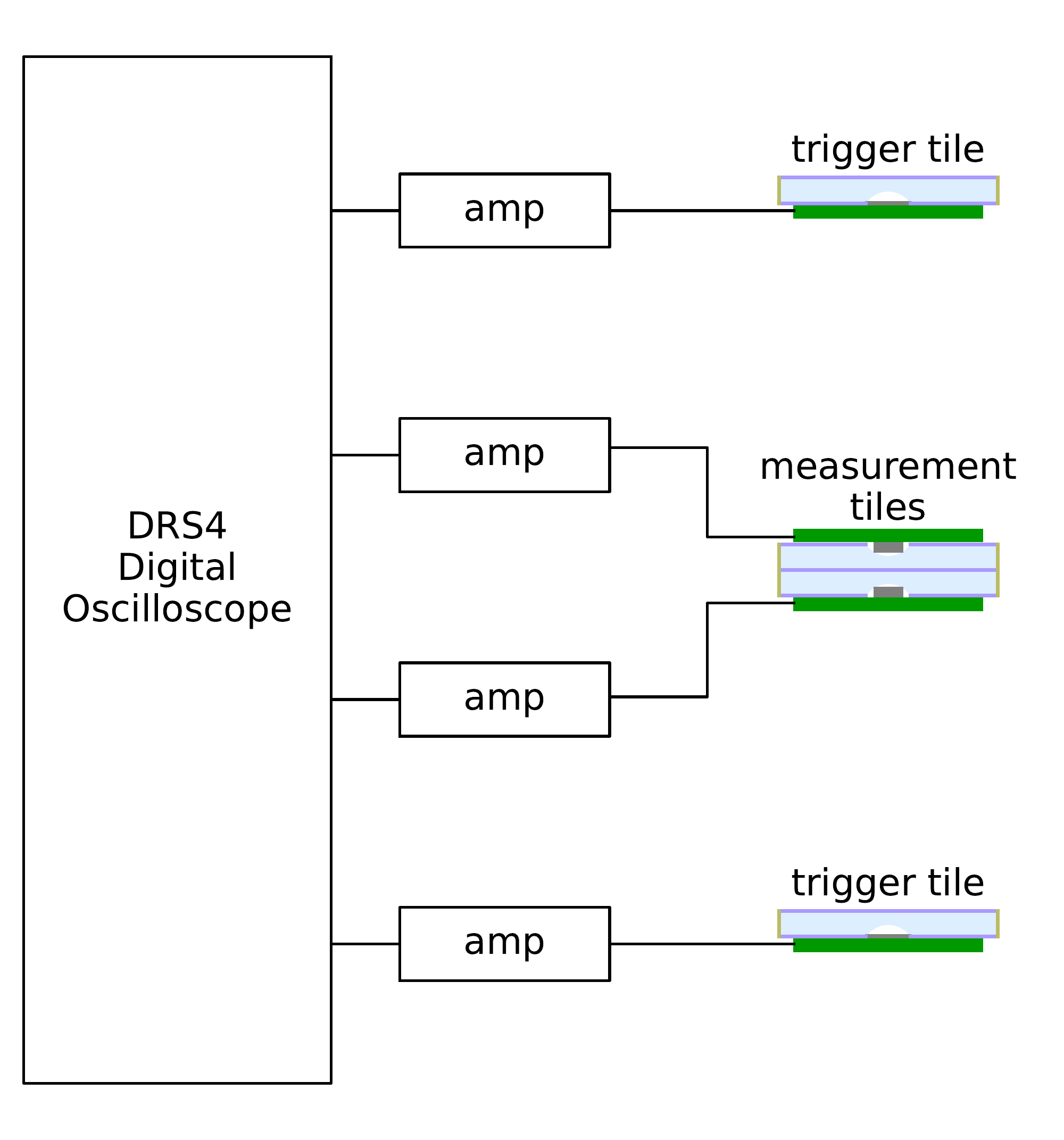}
\includegraphics[width=0.49\textwidth,trim={0 2cm 2cm 0},clip]{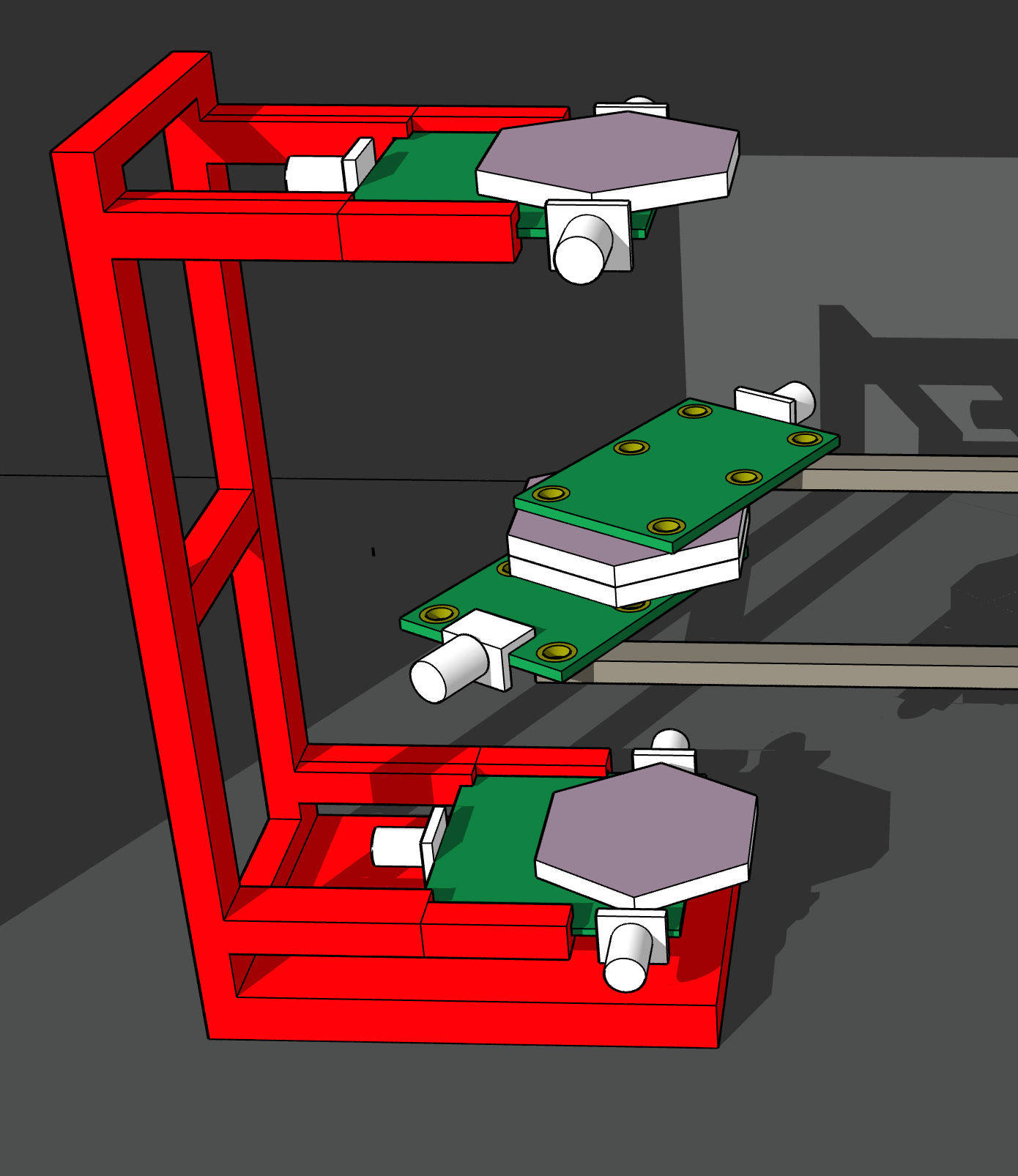}
\caption{\label{fig:cosmic_setup_timing} Block diagram (left) and a 3D rendering (right) of the cosmic-ray setup used for timing and light-yield measurements.}
\end{figure}

We used a three and four-fold coincidence setup to measure the signals from cosmic-ray muons.  
Two hexagonal tiles, each attached to 6 mm$^2$ SiPMs, were aligned with each other 10~cm apart vertically to act as the trigger. We placed two hexagonal tiles with SiPMs at the halfway point between the two trigger tiles, pressed back-to-back with each other, as shown in Fig.~\ref{fig:cosmic_setup_timing}.  

This setup allowed a measurement of the light yield from minimum-ionizing particles (MIPs), as well as the timing resolution.
For the timing-resolution measurement, we ensured that the cable distance between the SiPMs and the DRS4 were identical.  

In this setup, as well as the other two setups, the scintillator tiles, the SiPM-carrying boards, and their support structure were placed inside a dark box in order to prevent light leakage into the scintillators from outside light sources.  

\subsection{LED setup}
\begin{figure}[h!]
\centering 
\includegraphics[width=0.49\textwidth,clip]{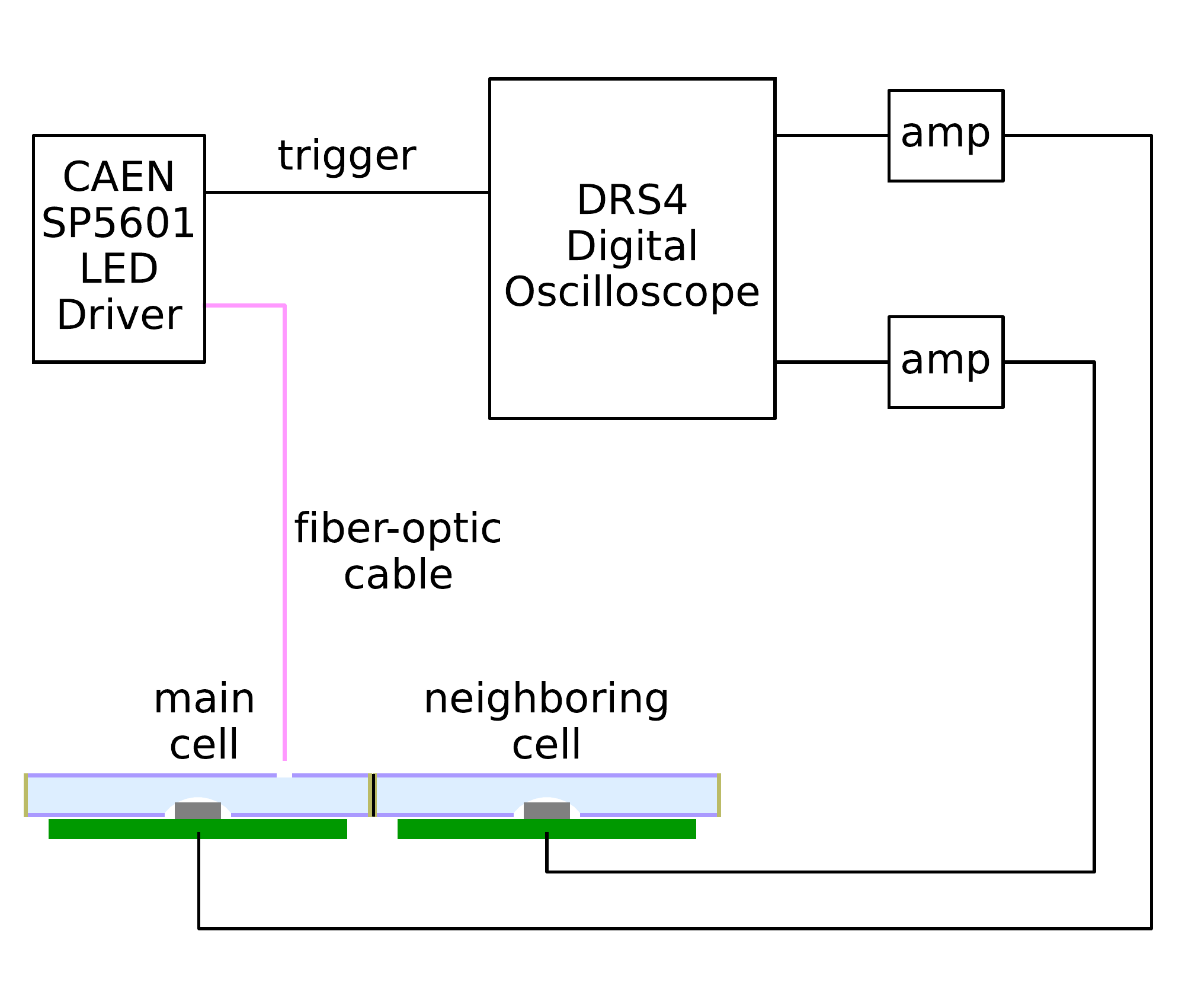}
\includegraphics[width=0.49\textwidth,clip]{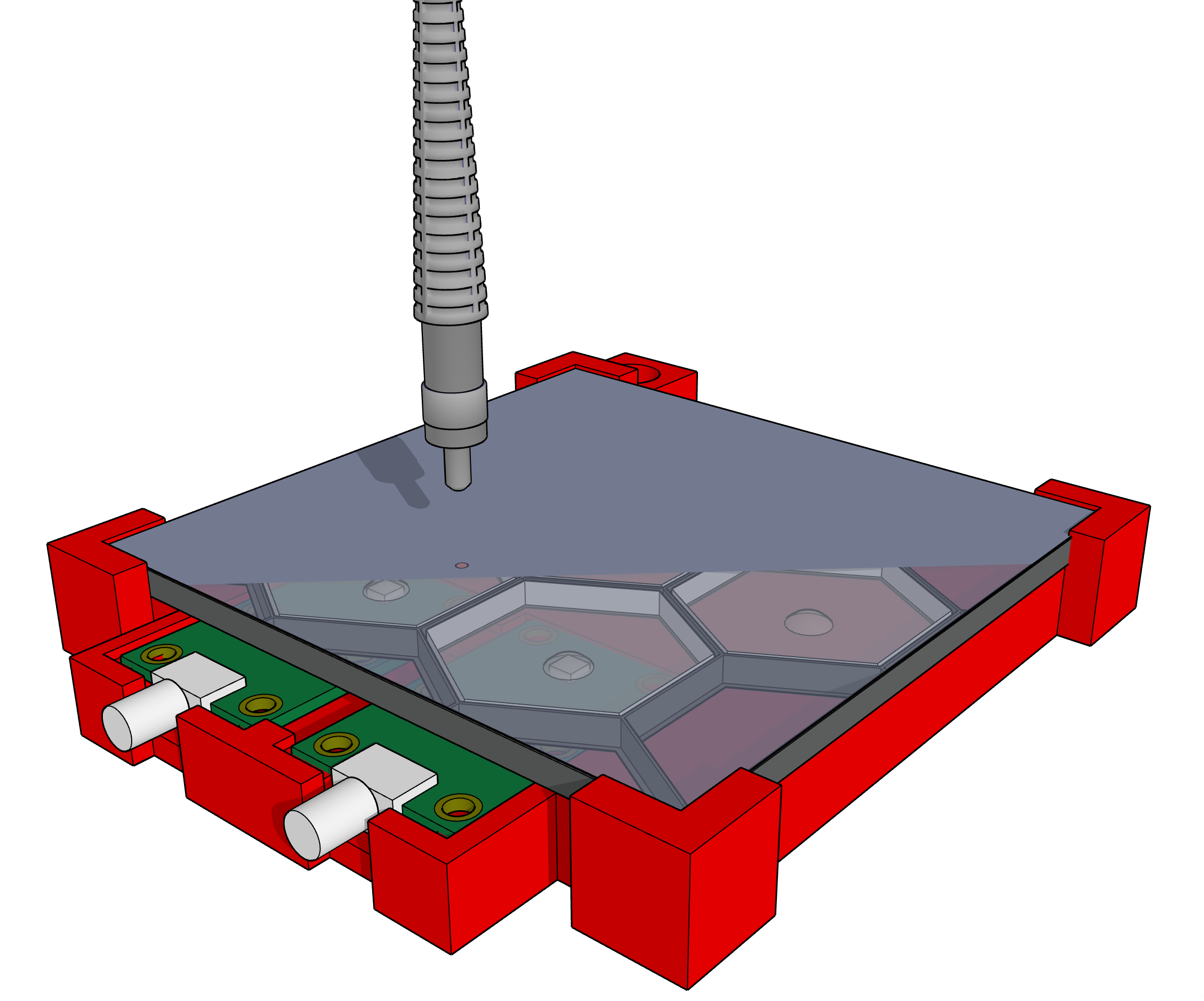}
\caption{\label{fig:led_setup} Block diagram (left) and a 3D rendering (right) of the UV-LED setup.  In the latter, the reflective foil is rendered as translucent, in order to show the 3D-printed frame, tiles, SiPMs and boards underneath.}
\end{figure}

To measure the level of optical crosstalk\footnote{This is not to be confused with SiPM crosstalk.}  between the scintillator cells, we developed the following system using a UV LED as our source.  The UV light was generated from a CAEN SP5601 LED Driver, and fed through a UV fiber-optic cable. The end of the cable was placed directly above either one of the cells in the megatile or one of the tiles in a 3D-printed frame.  A SiPM was placed under this cell (called the ``main'' tile) and another was placed under a neighboring cell. A pinhole in the reflective foil directly under the end of the fiber-optic cable allowed light to enter the tile. Both the SiPM output and the LED driver trigger signal were connected to a DRS4 digitizer connected to a computer. This setup is illustrated in Fig.~\ref{fig:led_setup}.  

In order to determine the level of optical crosstalk we manually adjusted the LED intensity, in order to sweep from zero to $\approx$150 photoelectrons in the main cell, at which point the limit of the DRS4 dynamic range was reached.  We define the level of optical crosstalk as the slope of the correlation between the number of photoelectrons per event in the neighboring cell and that of the main cell.  

\begin{figure*}
    \centering
    \includegraphics[width=0.9\textwidth, trim={0 5cm 0 0},clip]{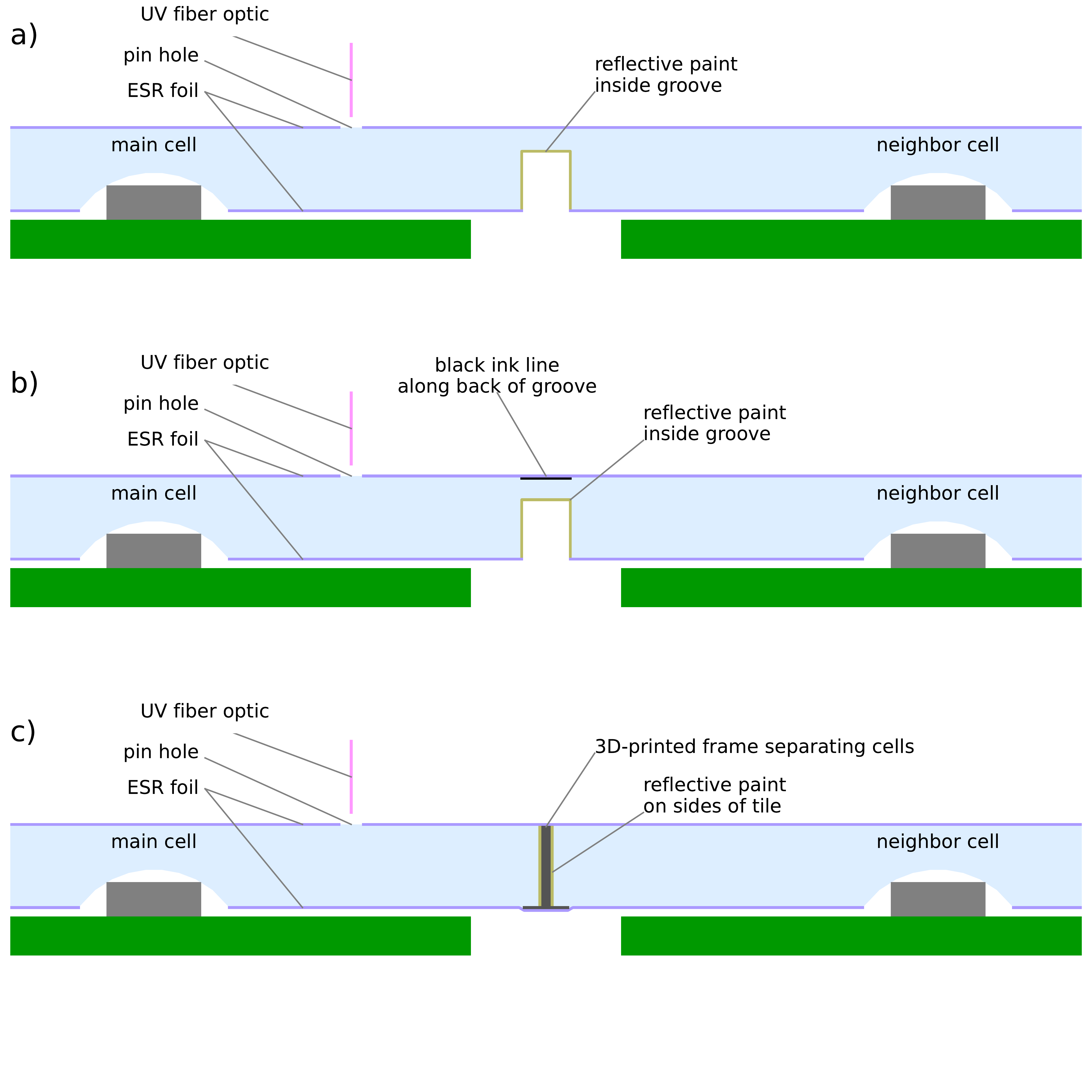}
    \caption{Comparison of the setups for testing methods of optical crosstalk reduction.  From top to bottom:  a) reflective paint inside grooves b) same as the previous, but with black ink on the back side of the scintillator opposite the grooves c) a 3D-printed frame used to separate individual cells.}
    \label{fig:crosstalk_setups}
\end{figure*}

We tested three different methods of reducing the optical crosstalk between the cells.  In the first method, the inside of the grooves were painted with reflective paint.  This would reflect the light that would otherwise enter the neighboring cell back into the main cell.  In the second method, the reflective paint was also used, and the face of the megatile opposite the grooves was traced with black ink in order to reduce reflection into the neighboring cell.  In the third method, the megatile was replaced with individual scintillator tiles inserted into a 3D-printed frame (see right panel of Fig.~\ref{fig:scintillators}) made of black, opaque PLA.  Since the walls between the tiles in the 3D-printed frame could be made thinner than the width of a groove in a megatile, the use of a 3D-printed frame has the potential to reduce the amount of deadzone between cells.  A cross-sectional schematic of the boundary between cells using each of these three methods is shown in Fig.~\ref{fig:crosstalk_setups}.

\subsection{Radioactive-Source Setup}

\begin{figure}
    \centering
    \includegraphics[width=0.45\textwidth]{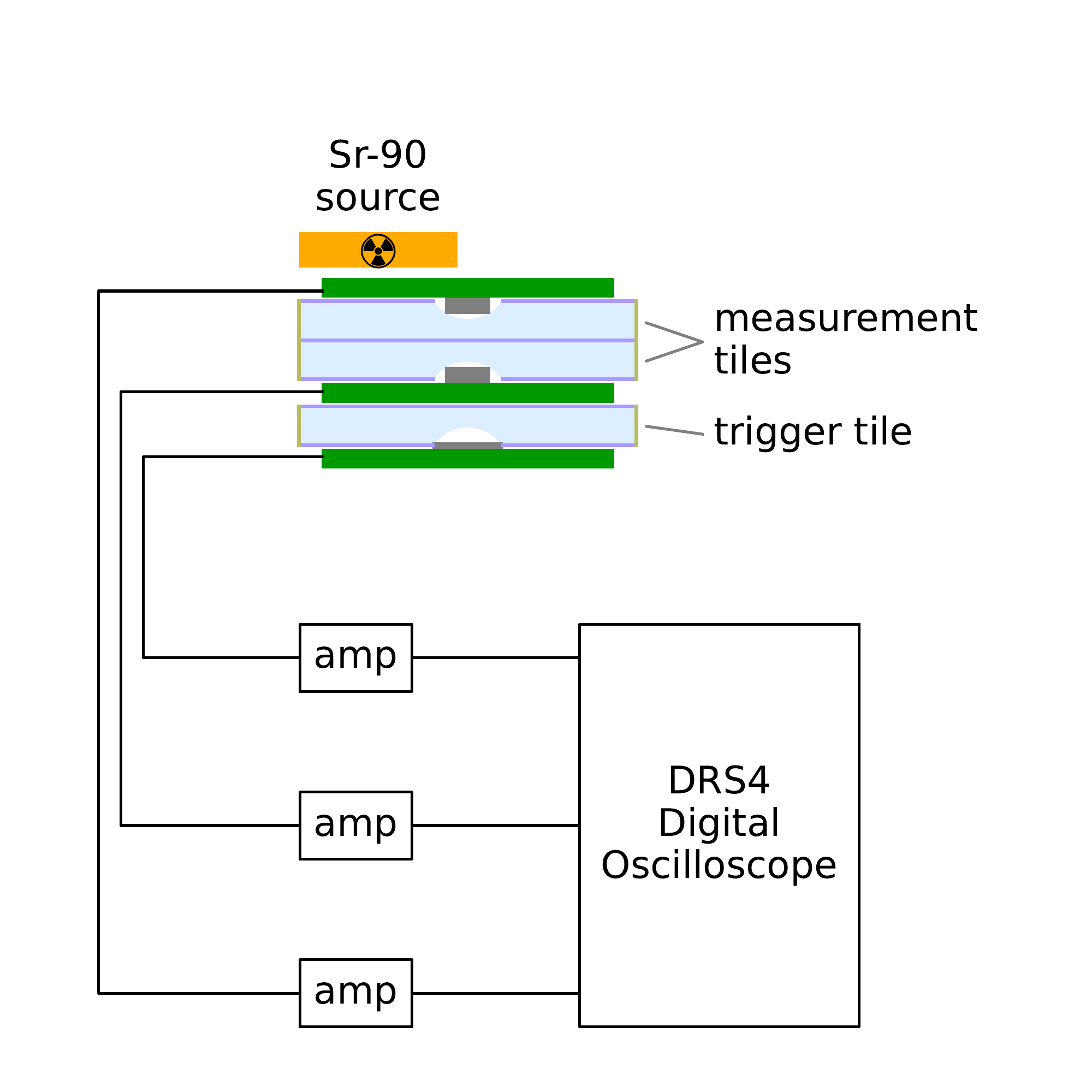}\includegraphics[width=0.45\textwidth]{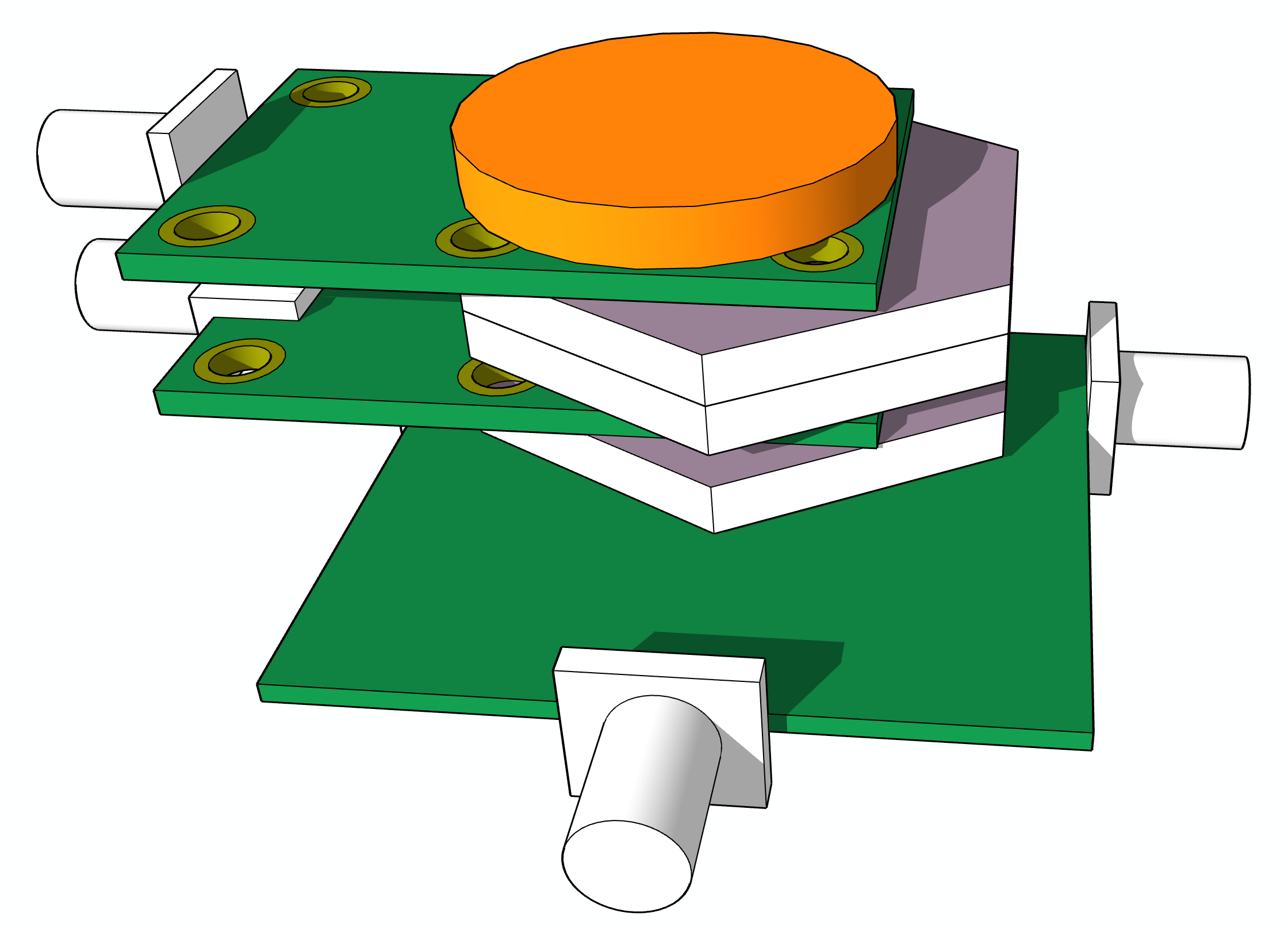}
    \caption{Setup for the tests with the radioactive source.  Left: block diagram.  Right: 3D rendering.}
    \label{fig:radioactive_setup}
\end{figure}

In the radioactive-source setup, we used an Sr-90 radioactive source in a sealed disk (SN-9796, 0.1~$\mu$Ci) to generate signals within the tiles.  Similar to the cosmic-ray setup, we used two back-to-back hexagonal tiles on the measurement boards for measuring the time and light yields. For the trigger, we used a third tile which was read out from one of the trigger boards, and placed directly below the lower of the two measurement boards.  This setup is shown in Fig.~\ref{fig:radioactive_setup}.  

\section{Results}
\subsection{Light Yield}
\begin{figure}[h!]
\centering 
\includegraphics[width=0.45\textwidth]{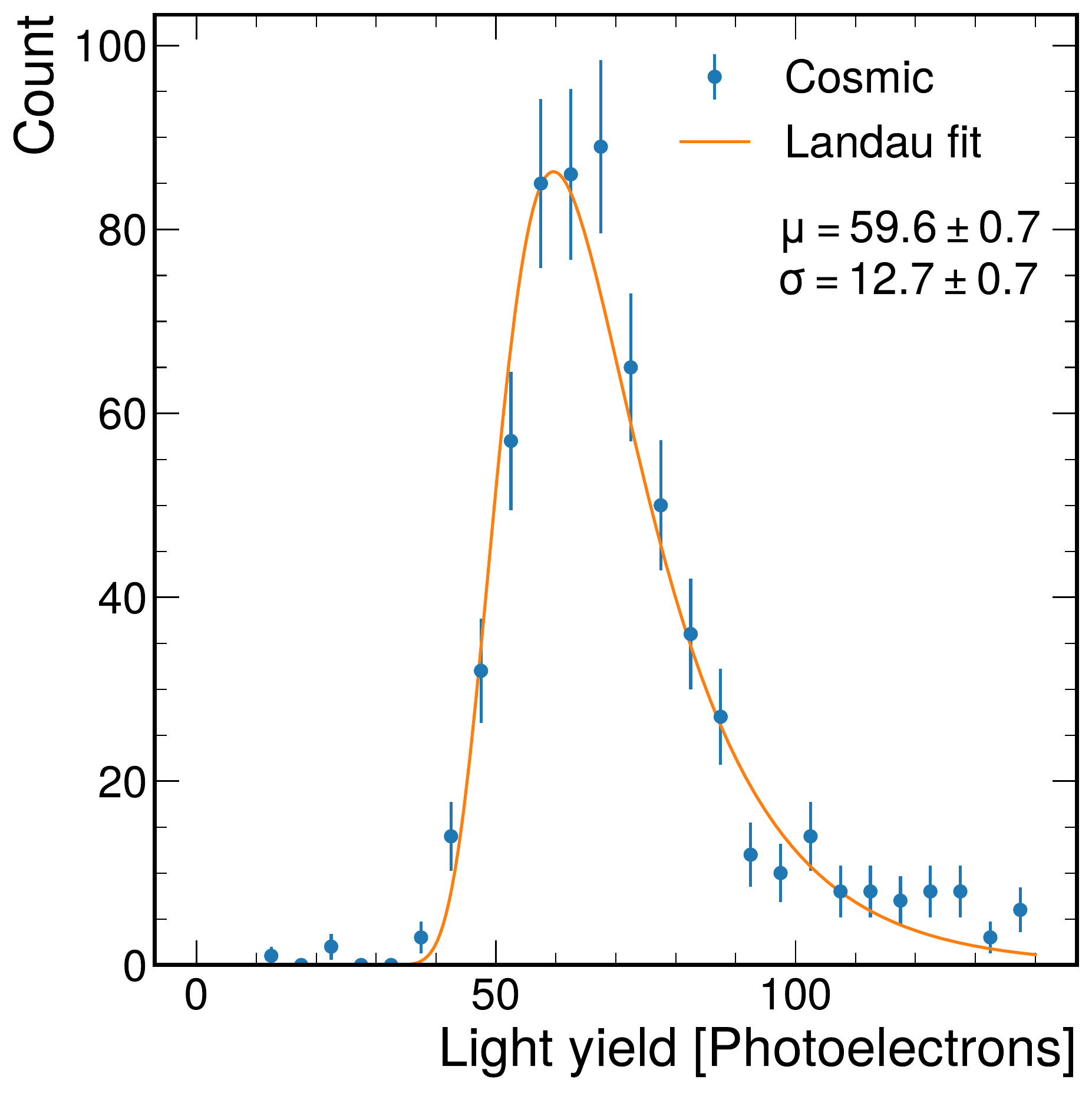}\includegraphics[width=0.45\textwidth]{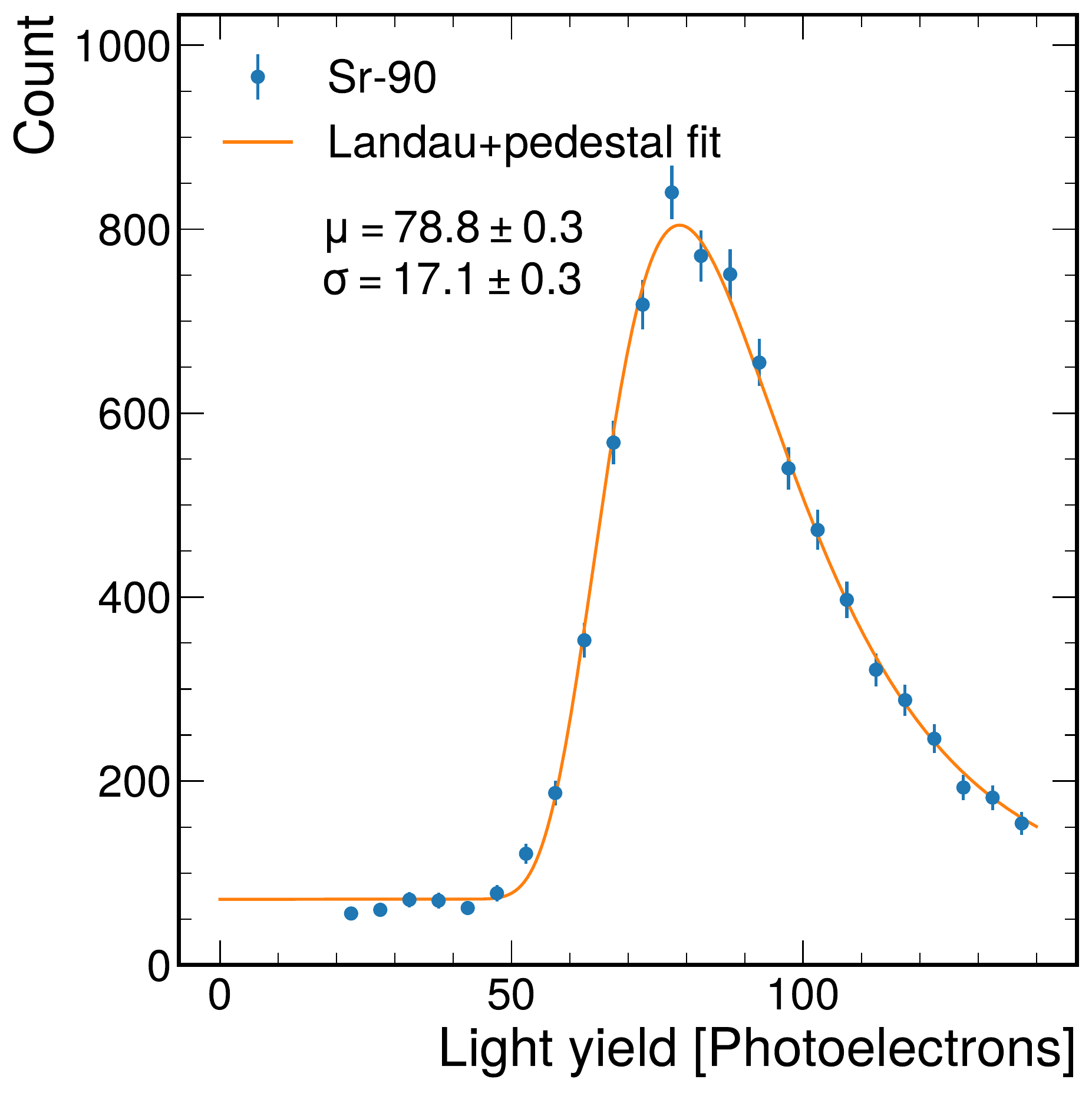}
\qquad
\caption{\label{fig:lightyields} Light yield measured with cosmic-rays (left) and Sr-90 (right).}
\end{figure}

The light-yield distribution for the SiPM on one of the measurement tile from the cosmic-ray setup is shown in the left panel of Fig.~\ref{fig:lightyields}.  We fit this distribution to a Landau function with most-probable value $\mu$, which is used do define the light yield. The results were $\mu=$59.6$\pm$0.7 and $\mu=$64.2$\pm$0.8 photoelectrons for the two SiPMs; the significant difference of about $\approx$8\% suggests a systematic uncertainty that could arise from the variations in tile optical quality or other source. 

As a cross-check, we determined the light-yield distribution for three-fold coincidence events with the Sr-90 setup (right panel of Fig.~\ref{fig:lightyields}) and fit it to a Landau function with an added constant term (to account for background\footnote{This background likely originates from accidental coincidences.}) and obtained $\mu$=78.8$\pm$0.3.  Strictly speaking, the electrons from Sr-90 are not MIPs, and the expected energy loss per unit length depends on their energy.  However, by requiring that the electrons produce 3-fold coincidence, our sample is biased towards the higher end of the possible phase space, where the maximum energy allowed in the Sr-90$\rightarrow$Y-90$\rightarrow$Zr-90 decay chain is 2.274 MeV.  In the 1.3-2.3 MeV range, the ratio of the energy loss per unit length for electrons to that of MIPs\footnote{As calculated using the formulas in Ref.~\cite{Groom:2000sm}.} is fairly flat, at around 1.3.  Dividing the $\mu$ from the Sr-90 by this ratio yields 61 photoelectrons per MIP, which is comparable to the value obtained with cosmic rays.   

Our light-yield results are larger than those reported in Ref.~\cite{Belloni_2021}, which obtained about 32 photoelectrons per MIP for a similarly sized square cell ($3\times 3$~cm$^2$), using $1.3\times1.3$~mm$^2$ SiPMs from the same product series as we used, and in Ref.~\cite{Jiang:2020tve}, which likewise obtained a value of 28 photoelectrons per MIP in a similar setup.  While our light yield was larger than those of Refs.~\cite{Belloni_2021,Jiang:2020tve}, this improvement likely originates from the larger surface area of the SiPMs used, although it does not scale linearly with SiPM size.

\subsection{Optical crosstalk between channels}
\begin{figure}[h!]
    \centering
    \includegraphics[width=0.99\textwidth]{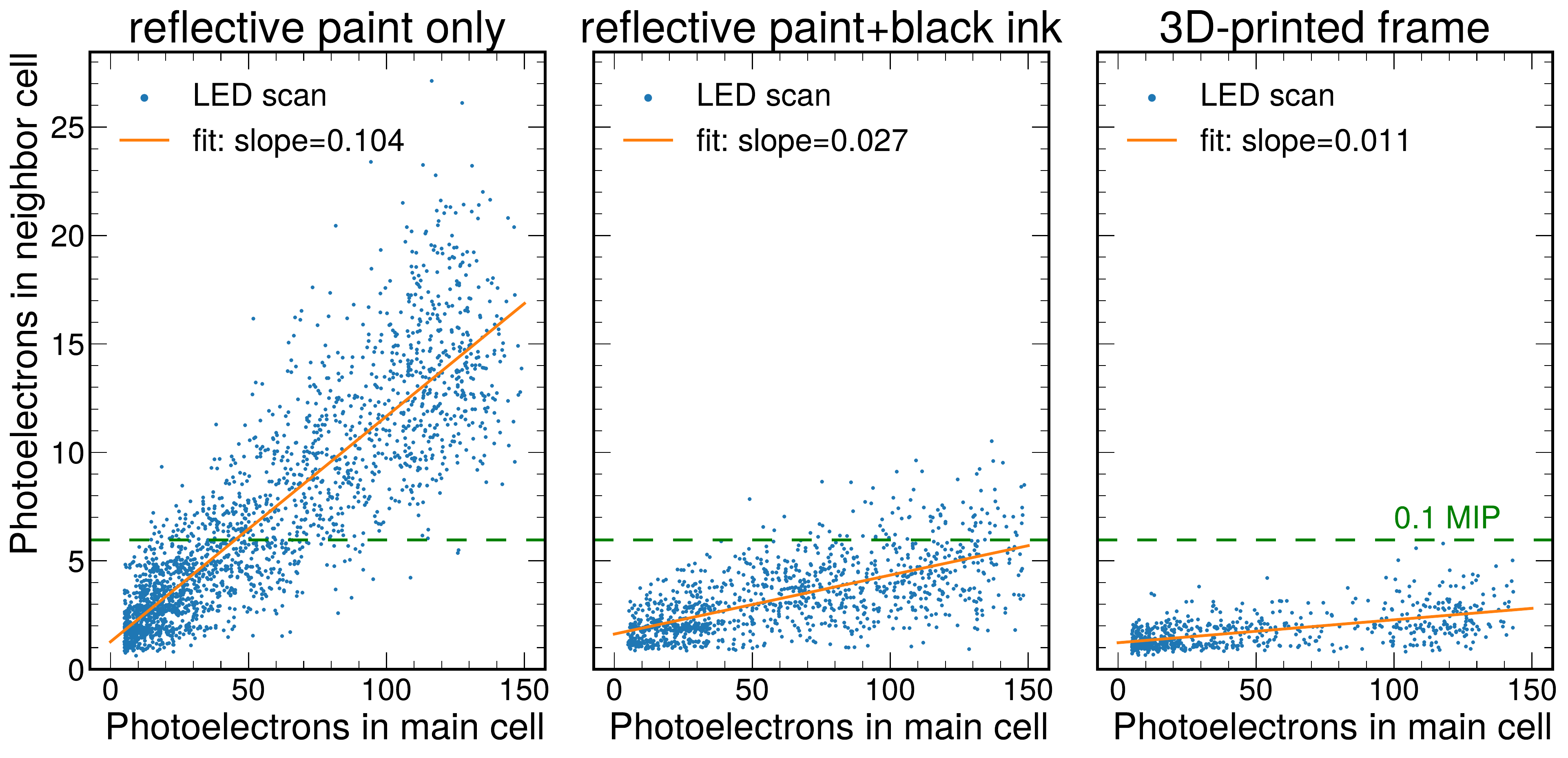}
    \caption{Comparison between the amounts of crosstalk between neighboring cells, using three methods of reduction:  painting inside the grooves with reflective paint (left), using the reflective paint and also tracing the face opposite the grooves with black ink (middle), and fitting the scintillators into a 3D-printed frame.  The dashed green line represents 0.1 times the number of photoelectrons expected from a MIP.}
    \label{fig:crosstalk}
\end{figure}

We used the UV-LED setup for the evaluation of the optical crosstalk between cells.   To compare the effectiveness of each of the three methods of optical crosstalk reduction, we plotted the signal in the main cell against the signal in the neighboring cell in Fig.~\ref{fig:crosstalk}. 

 Due to limitations of the dynamic range of the digitizer, we only include data with up to 150 photoelectrons in the main cell, corresponding to 2.5 MIPs. 
 We used a linear regression on these data, and determined the slopes:  0.104 for the reflective paint only (left), 0.027 with the reflective paint plus black ink (middle), and 0.011 with the 3D-printed frame (right).  Overall, these results suggest that the 3D-printed frame is the most effective for reducing crosstalk. 

    Typically, one defines a threshold using some fixed fraction of the number of photoelectrons corresponding one MIP, \textit{e.g.} in Ref.~\cite{hcalInsert} this was assumed to be 0.1.  Therefore, we show in Fig.~\ref{fig:crosstalk} a line corresponding to 0.1 MIPs in the neighboring cell.  We find that for the reflective-paint-only megatile, this threshold is exceeded for most of $x$-axis range.  When adding black paint, the average signal in the neighbour cell never exceeds this threshold but some events do.  With the 3D-printed frame, the average is well below the threshold and none of the events in this sample cross it.  
    
    The detector response of the HG-CALI to a 50~GeV $\pi^-$ was simulated in Ref.~\cite{hcalInsert}, predicting a cell-energy average of 2.0 MIPs, which is within the range probed in this study.  However, some of the hits in the simulation had much larger signals (up to 250 MIPs, which defines the needed dynamic range per cell).  Extrapolating from our measurement,
    these would correspond to a neighboring-cell signal of about 2.8 MIPs; however, such high-energy hits were rare in the simulation.

As a cross-check, we used a variant of our cosmic-ray setup in which we placed the trigger tiles directly above and below the main cell of a megatile or 3D-printed frame.  The optical crosstalk would then be the ratio between the peak values of the distributions of the signals in the neighboring and main cells.  We repeated this for the megatile without ink, the megatile with ink, and the 3D-printed frame.  The optical crosstalk values were consistent with those obtained using the UV-LED setup; in the latter case, the measured distribution is consistent with noise. 

\subsection{Timing resolution}
\begin{figure}[h!]
\centering 
\includegraphics[width=0.45\textwidth,clip]{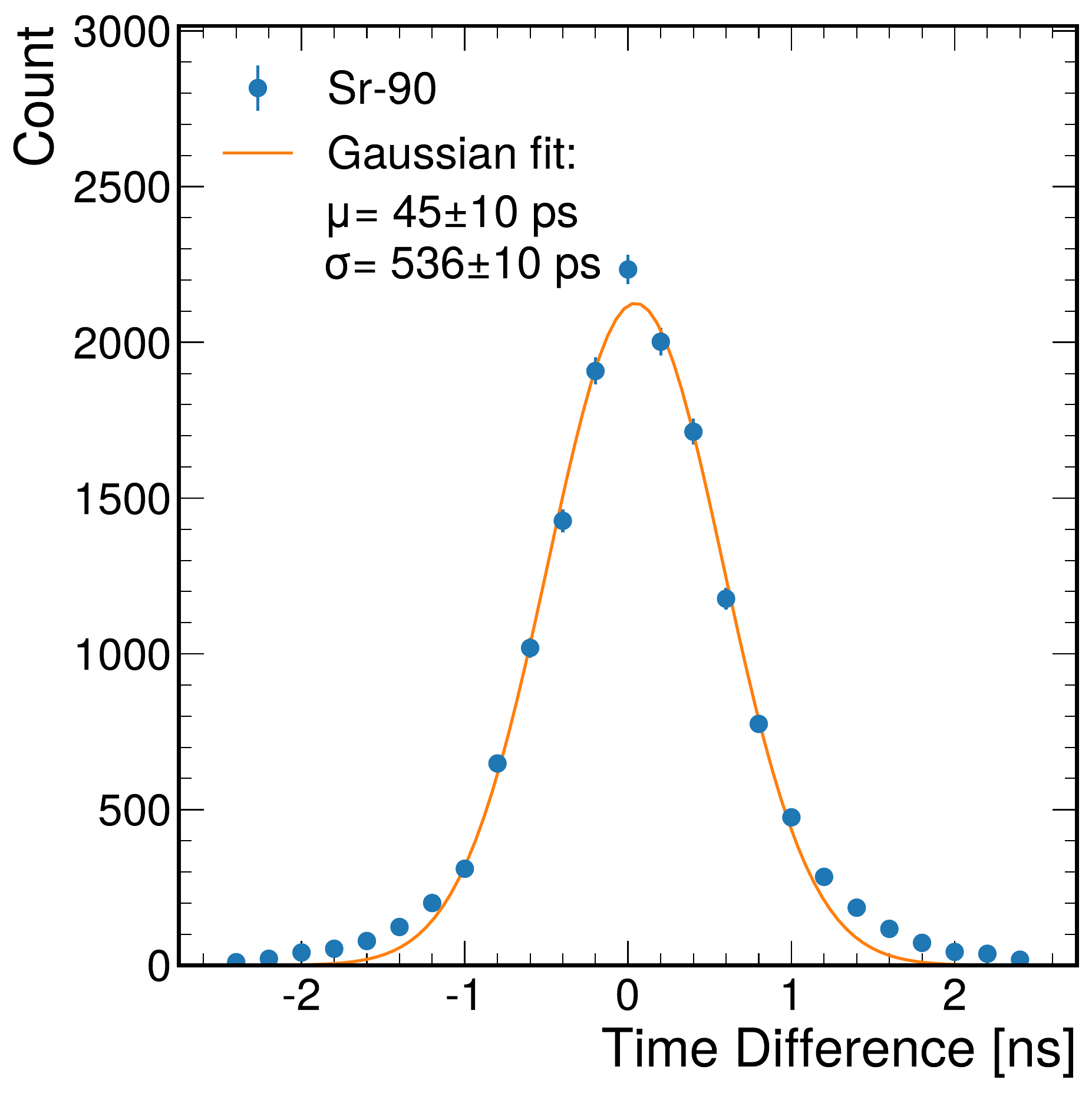}\includegraphics[width=0.45\textwidth,clip]{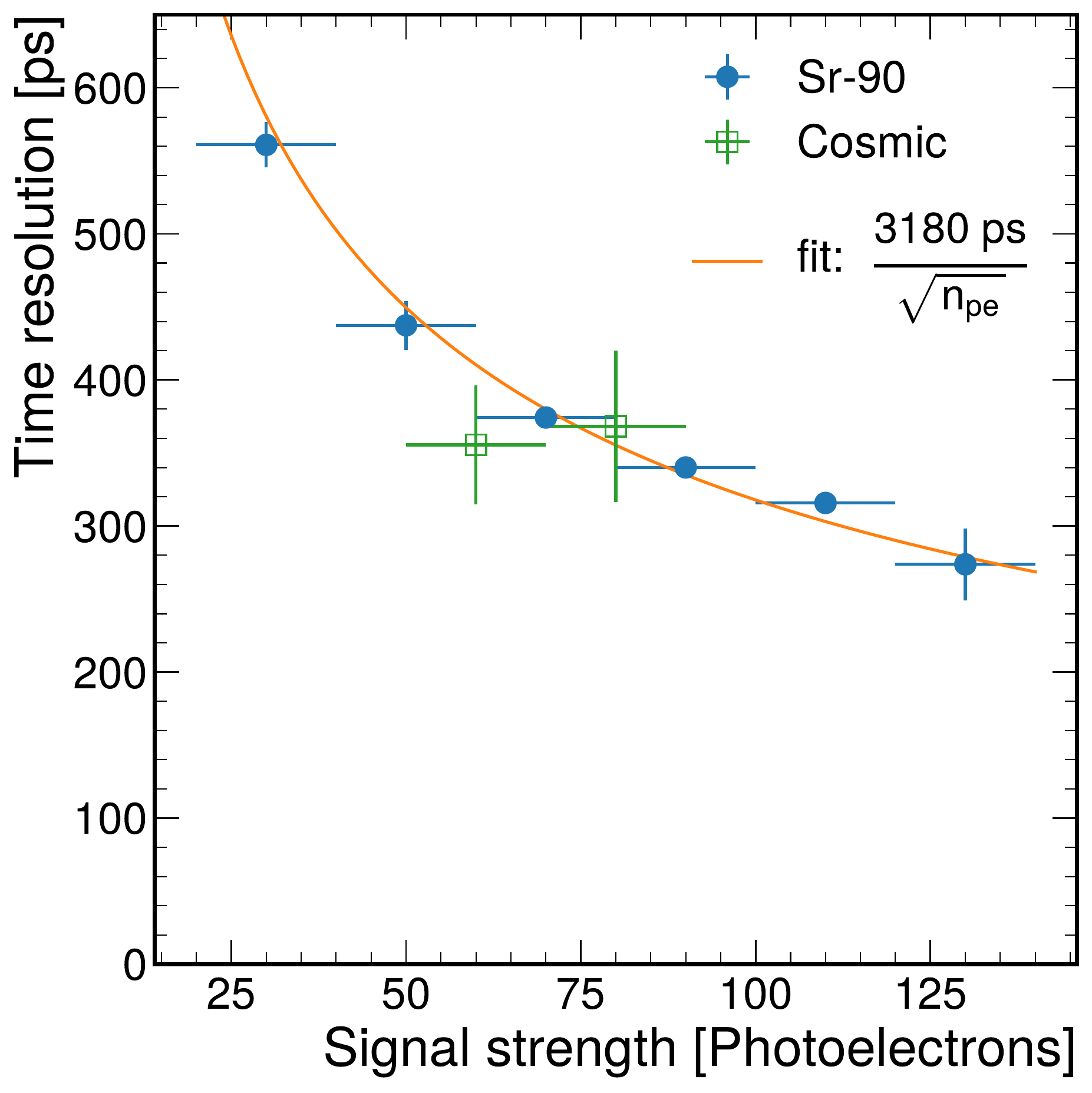}
\caption{\label{fig:time_results} Left: Time difference of two SiPMs using an Sr-90 source. A Gaussian fit is shown as an orange curve.  Right: Time resolution as a function of the number of photoelectrons obtained using an Sr-90 source (blue circles) and cosmic rays (green squares). A fit to a constant times $1/\sqrt{n_{pe}}$ is shown as an orange curve.}
\end{figure}

We show the distribution of the time differences between the two back-to-back SiPMs in the Sr-90 timing setup in the left panel of Fig.~\ref{fig:time_results}.  We fit this to a Gaussian function, yielding $\mu=$46$\pm$9~ps and $\sigma=$540~ps.  To obtain the resolution for a single SiPM, we divided the $\sigma$ of this Gaussian by the square root of two, yielding a single-SiPM timing resolution of 380~ps. This is comparable to the 0.5 ns resolution reported in Ref.~\cite{Chatrchyan_2018} for plastic
scintillator tiles at the CERN H2 test beam.

We repeated this process while simultaneously binning the data in the number of photoelectrons in the two SiPMs in the same bins, and show the result in the right panel of Fig.~\ref{fig:time_results}, both for the Sr-90 setup (blue circles) and for the cosmic-ray setup (green squares)\footnote{In the left panel of Fig.~\ref{fig:time_results}, the requirement of having similar light yields in each cell was not used.}.  We observe that the timing resolution fits to $\sigma=$3180/$\sqrt{n_{\rm pe}}$, where $n_{\rm pe}$ is the number of photoelectrons.  This is slightly better than the preliminary results reported in Ref.~\cite{hummer_emberger_simon} for similarly sized cells. We repeated this measurement using all our algorithms to extract the time-of-arrival of signals; the results varied by only about 2\% from algorithm to algorithm.  

It is often the case that the timing of a signal will have an amplitude dependence, known as a ``time-walk'' effect, and this can depend largely on the algorithm used for defining the signal time.  We investigated this effect and found that this change in the time-of-arrival can vary by about 250~ps when one SiPM has a very large signal and the other has a very small signal.  However, this effect has a negligible impact on the resolution shown in the right panel of Fig.~\ref{fig:time_results}, because we required the two channels to have similar light yields, causing the time-walk effects on the two channels to cancel each other out.  

We note that both the timing resolution and the size of the time-walk corrections reported here are much smaller than those reported in Ref.~\cite{CALICE:2019ipm}, though this is likely due the fact that they used waveshifting fibers (which we didn't use) and slower readout.

Timing resolutions of less than 100~ps have been achieved in SiPM-on-tile approach in Refs.~\cite{6898046,Torres:2019ntb,Alvarado:2019cam}, though the resolution for a given setup may depend on many factors including the SiPM type and the material and geometry of the tile. 

It has been proposed in Ref.~\cite{Graf_2022} that using timing information to identify the presence of slow neutrons in showers can be used to improve shower-energy reconstruction, using advanced algorithms such as those in \cite{Akchurin:2021afn,Qasim:2021hex}.  Having good resolution in timing per cell can therefore be useful in such energy corrections.

\section{Summary and conclusions}
We performed studies with scintillating tiles read-out with SiPMs as part of the R\&D for the high-granularity calorimeter insert at the EIC.  In these studies, we have developed and tested two new innovations for the SiPM-on-tile approach that can simplify the assembly process of these tiles and improve their performance.

The first innovation was the use of opaque 3D-printed frames to insert the hexagonal scintillator cells into, which is an alternative to using grooved megatiles.  We found that the optical crosstalk was significantly lower ($\approx$1\%) with the 3D-printed frame than with the megatile ($\approx$3\%), even when black ink was traced opposite the grooves of the megatile.  Further, the use of these frames simplified the production process and reduced the dead zones between cells.  

The second innovation exploited the 3D-printed frame by using reflective paint on the edges of the cells and only using foil on the top and bottom. This approach simplifies the assembly of individual tiles and is an alternative to the complicated task of fully wrapping the scintillators in foil.  Further, the painted-edges method has the potential to reduce cell-by-cell variations which can occur for foil-wrapped scintillators. 

To further test the viability of using this combination of flat reflective foils and reflective paint, we performed measurements of the light yield and timing resolution for these cells.  We found that a MIP signal yields about 60 photoelectrons per cell using a 3$\times3$ mm$^{2}$ SiPM at 2 V overvoltage, which would be adequate for the EIC calorimeter insert~\cite{hcalInsert}. We also found an acceptable timing resolution of 600~ps at 0.5~MIP to 300~ps at about 2~MIP.  

Further studies can be done to optimize the light yield and uniformity, for instance by repeating these  and other studies with various dimple geometries, as guided by earlier optimization studies such as \cite{Belloni_2021,de_Silva_2020,Chadeeva_Korpachev_Rusinov_Tarkovskii_2018,Simon:2010hf,BOBCHENKO2015166,Jiang:2020tve,ABUAJAMIEH2011348,Liu:2015cpe,Blazey:2009zz,Simon:2010hf}.  
More efficient SiPMs, such as those that use epitaxial-quenching resistors~\cite{NDL,instruments1010005} may further improve light yield as demonstrated in Ref.~\cite{Jiang:2020tve}.
Additionally, cells made with the injection molding technique \cite{instruments1010005} could help improve cell uniformity and simplify the assembly process even further. Alternatively, the scintillating cells along with the opaque frame and reflective material could be produced as a single piece with a 3D-printing technique~\cite{Berns_2022,Berns:2020ehg} that could facilitate the production of the irregular cells on the edge of the detector. Future reconstruction studies involving multiple showers will be necessary to quantify the physics impact of the cross-talk reduction that we report in this paper. 

Overall, these studies demonstrate the performance of the building blocks of the high-granularity calorimeter insert for the EIC~\cite{hcalInsert} and show advancements in the SiPM-on-tile technology that could also inform other EIC calorimeters and other experiments.

\appendix

\acknowledgments
We thank members of the California EIC consortium, and in particular Oleg Tsai, for valuable feedback related to our design and studies. We thank Sergey Los and James Freeman for providing us with the SiPM boards that we used in this work. We also thank Martin Purschke for helping us with the RCDAQ software. We thank Ron Soltz and Ernst Sichtermann for supporting and guiding Luis Garabito Ruiz, Jiajun Huang, and Miguel Rodriguez. 

This work was supported by MRPI program of the University of California Office of the President, award number 00010100. This material is based upon work supported by the U.S.~Department of Energy, Office of Science, Office of Nuclear Physics, RENEW under Award Number DE-SC0022526, which supported Luis Garabito Ruiz, Jiajun Huang and Miguel Rodriguez. Sebouh J.~Paul acknowledges support by the Jefferson Laboratory EIC Center Fellowship. Sean Preins was supported by a HEPCAT fellowship from DOE award DE-SC0022313. Miguel Arratia acknowledges support through DOE Contract No. DE-AC05-06OR23177 under which Jefferson Science Associates, LLC operates the Thomas Jefferson National Accelerator Facility.

\bibliographystyle{utphys} 
\bibliography{biblio.bib}

\end{document}